\documentclass[a4paper,fleqn,usenatbib]{mnras}

\usepackage[T1]{fontenc}
\usepackage{ae,aecompl}

\usepackage{graphicx}	
\usepackage{amsmath}	
\usepackage{amssymb}	
\usepackage{hyperref}

\newcommand{\cmsq}{\mbox{cm$^{-2}$}}
\newcommand{\nc}{\newcommand}
\nc{\kms}{\mbox{km~s$^{-1}$}}
\nc{\cmcub}{\mbox{cm$^{-3}$}}
\nc{\av}{\mbox{$A_{\rm V}$}}

\title[OH in diffuse clouds]{Revisiting the OH-CH correlation in diffuse
clouds}

\author[B. Mookerjea et al.]{
Bhaswati Mookerjea\thanks{E-mail: bhaswati@tifr.res.in}
\\
$^{1}$Tata Institute of Fundamental Research, Homi Bhabha Road, Mumbai
400005, India\\
}

\date{Accepted 2016 April 12. Received 2016 April 08; in original form
2016 February 26}

\pubyear{2016}

\begin{document}
\label{firstpage}
\pagerange{\pageref{firstpage}--\pageref{lastpage}}
\maketitle

\begin{abstract}

Based on the  analysis of available published data and archival
data along 24 sightlines (5 of which are new) we derive more accurate
estimates of the column densities of OH and CH towards
diffuse/translucent clouds and revisit the typically observed
correlation between the abundances of these species. The increase in
the sample size was possible because of the equivalence of the column
densities of CH derived from a combination of the transitions at 3137
\& 3143\,$\AA$,  and a combination of transitions at 3886 \& 3890\,$\AA$,
which we have demonstrated here.  We find that with the exception of
four diffuse clouds, the entire source sample shows a clear correlation
between  the column densities of OH and CH similar to previous
observations.  The analysis presented also verifies the theoretically
predicted oscillator strengths of the OH A--X (3078 \& 3082\,$\AA$), CH
B--X (3886 \& 3890\,$\AA$) and C--X (3137 \& 3143$\AA$) transitions. We
estimate $N$(H) and $N$(H$_2$) from the observed $E(B-V)$ and $N$(CH)
respectively. The $N$(OH)/$N$(CH) ratio is not correlated with the
molecular fraction of hydrogen in the diffuse/translucent clouds.  We
show that with the exception of HD\,34078 for all the clouds the
observed column density ratios of CH and OH can be reproduced by simple
chemical models which include gas-grain interaction and gas-phase
chemistry. The enhanced $N$(OH)/$N$(CH) ratio seen towards the 3 new
sightlines can be reproduced primarily by considering different cosmic
ray ionization rates.

\end{abstract}




\section{Introduction}

The OH molecule was discovered using the $\Lambda$ doublet transition
observed between the levels of the ground rotational state
$^{2}\Pi_{3/2}$ J = 3/2 at 18 cm \citep{weinreb1963}; later its
electronic transitions were identified in ultraviolet spectra of
bright OB-stars \citep{crutcher1976,chaffee1977, felenbok1996}. Two
lines of the A$^2\Sigma^+$--X$^2\Pi_i$ band near 3078 and 3082 $\AA$
are available to ground-based observatories and in a series of papers
\citep{weselak2009,weselak2010} estimates of OH column densities along
17 translucent lines of sight from these transitions have been
derived.  These observations performed using the high-resolution
VLT/UVES-spectrograph verified the oscillator strengths for the 3078
and 3082\,$\AA$ and also suggested a close correlation between the
column densities of OH and CH molecules. Early chemical models have
suggested production of hydrides like OH and NH due to grain catalysis
reactions \citep{wagenblast1993}. Hence observation of OH to derive
abundances towards translucent sightlines with larger than normal
far-UV extinction is interesting.  Interstellar CH is often observed
using its strongest A--X transition at 4300$\AA$.  However this line
being saturated, reliable estimates of CH column densities are derived
from the spectral lines due to the weaker B--X system at 3886$\AA$
\citep{krelowski1999,weselak2004}. There exist many studies of CH to
explore its correlation to the diffuse interstellar bands (DIBs) at
5780 and 5797$\AA$ \citep{krelowski1999}.  However availability of
data for OH is significantly worse. Observations have shown the
abundances of CH molecule to be tightly correlated with those of the
H$_2$ molecule \citep{mattila1986,weselak2004,sheffer2008}. The
correlation is sufficiently strong so that in many recent studies
column densities of CH have been used to derive estimates for H$_2$
column density \citep{sheffer2008,wiesemeyer2016}. The correlation is
understood in terms of higher formation rates of CH in reactions
involving molecular hydrogen.

The aim of this work is to incorporate the newly available OH and CH
data for 13 lines of sight (5 of which are new) in order to
investigate the correlation between OH and CH in diffuse/translucent
molecular clouds. This correlation is particularly interesting
considering the tight correlation between CH and H$_2$. We shall show
that the oscillator strengths for the C--X system of lines of CH at
3137\,$\AA$  are correct so that these can be equivalently used to
determine the column densities of CH.  We shall use simple chemical
models to check whether the observed correlation between OH and CH or
the absence of it is predicted by the chemical models.

\section{UV Spectroscopic datasets}

We extend the analysed sample of \citet{weselak2010} by combining it
with  sample of the highly reddened early-type stars which were
analyzed by \citet{bhatt2015}. In order to complete the information on
all the lines of sight as much as possible, we have also analyzed fully
processed UVES/VLT data of both pairs of CH lines, obtained from the
ESO Science Archive
Facility\footnote{\url{http://www.eso.org/sci/observing/phase3.html}}.
\citet{bhatt2015} had identified 30 known features (11 atomic and 19
molecular) and tentatively detected up to 7 new interstellar absorption
lines of unknown origin towards 346 targets. Out of this entire sample
only for 13 sources observations of both the 3078 and 3082\,$\AA$
transitions of OH were available. Out of these 13 sources 8 were
already present in the sample used by \citet{weselak2010}.  The
observed data for the two CH transitions B$^2\Sigma^-$--X$^2\Pi_i$ at
3886 and 3890\,$\AA$ for 19 of these sightlines were also presented by
\citet{weselak2010}.  \citet{bhatt2015} presented the
C$^2\Sigma^+$--X$^2\Pi_i$ transitions of CH at 3137 and 3143\,$\AA$ for
the 12 out of the 13 sources in which OH was observed.  Thus we finally
have 24 sources, 19 from \citet{weselak2010} and 13 from
\citet{bhatt2015}, of which 5 sightlines are new.

\section{Results and discussion}

Table\,\ref{tab_spectro} presents the spectroscopic details,
frequencies, and $f$-values of the transitions of OH and CH considered
here.

\begin{table*}
\caption{Spectroscopic details and $f$-values of OH and CH transitions studied.
CH transitions 
\label{tab_spectro}}
\begin{tabular}{crclccc}
\hline
\hline
\multicolumn{1}{c}{Species} & 
\multicolumn{1}{c}{Vibrionic Band} & 
\multicolumn{1}{c}{Rotational lines} & 
\multicolumn{1}{c}{Position} & 
\multicolumn{1}{c}{Ref. }&
\multicolumn{1}{c}{$f$-value} &
\multicolumn{1}{c}{ Ref.}\\
\hline
\hline
OH  & A$^2\Sigma^+$--X$^2\Pi_i$ (0,0) & Q$_1$(3/2)+$^QP_{21}$(3/2) & 3078.443 & 1 & 0.00105 & 2\\
    &                           (0,0) & P$_1$(3/2) & 3081.6643& 1 & 0.000648 & 2\\
\hline
CH  & C$^2\Sigma^+$--X$^2\Pi_i$ (0,0) & R$_2$ & 3137.576 & 3 & 0.00210 & 3 \\
    &                           (0,0) & Q$_2$(1)+$^QR_{12}$(1) & 3143.150 & 3 & 0.00640 & 3 \\
CH  & B$^2\Sigma^-$--X$^2\Pi_i$ (0,0) & Q$_2$(1)+$^QR_{12}$(1) & 3886.409 & 3 & 0.00320 & 3 \\
    &                           (0,0) & $^PQ_{12}$(1) & 3890.217 & 3 & 0.00210 & 3 \\
\hline
\end{tabular}

1. Weselak et al. (2009), 2. Felenbok \& Roeuff (1996) and 3. Lien (1984)
\end{table*}   

Table\,\ref{tab_meas} presents the observed equivalent widths (EW) of
the OH lines at 3078 and 3082\,$\AA$ and Fig.\,1 shows a plot of the
same.  We find that excluding HD\,114213 (with the largest
EW(3082$\AA$)), the equivalent widths of the two transitions show a
correlation of 0.92, which is somewhat worse than the estimates of
\citet{weselak2009}. From the slope of the line of regression we
estimate the ratio of the equivalent widths of the two transitions to
be 1.62$\pm$0.08.  The wavelengths of the two transitions being
similar, the equivalent widths are expected to have a ratio similar to
the ratio of the $f$-values of the two transitions.  The derived ratio
matches well with the estimate (1.62) based on the $f$-values of the
two transitions.

\begin{figure}
\centering
\includegraphics[width=\columnwidth,angle=0]{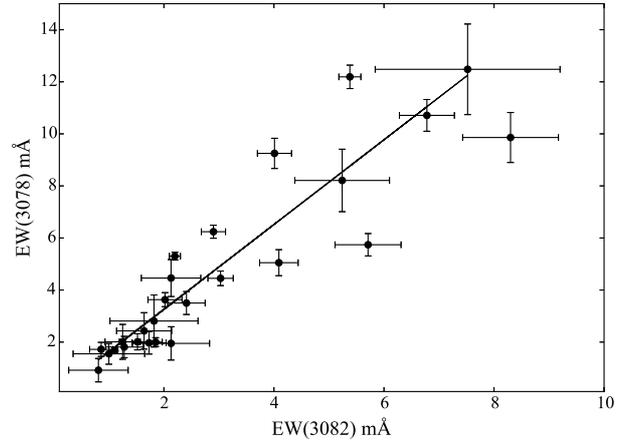}
\caption{Comparison of the equivalent widths of the two transitions of OH at 3078 and
3082\,$\AA$. The straight line corresponds to a linear regression with
a correlation coefficient of 0.92 and a slope of 1.62$\pm$0.08.}
\label{fig_ohcomp}
\end{figure}

For all the unsaturated lines of OH and CH we estimate the column
densities using the relationship proposed by \citet{herbig1968}:

\begin{equation}
N = 1.13\times 10^{20} W_\lambda/(\lambda^2 f)
\end{equation}

where $W_\lambda$ and $\lambda$ are in $\AA$,  column density is in
\cmsq\ and the $f$-values used are taken from
Table\,\ref{tab_spectro}. The total column density for OH is estimated
by adding the column densities of the 3078 and 3082\,$\AA$ transitions
(Table\,\ref{tab_meas}).  

\begin{figure}
\centering
\includegraphics[width=\columnwidth,angle=0]{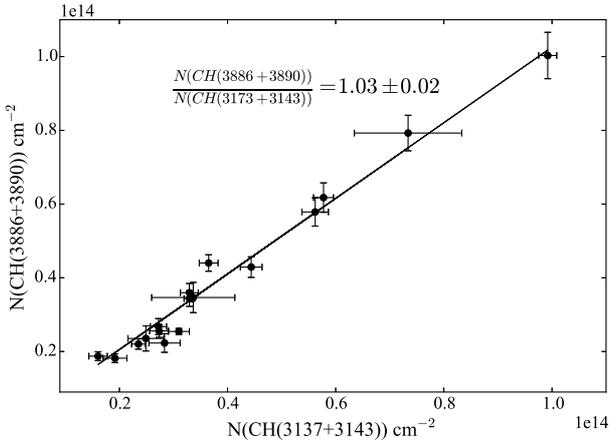}
\caption{Comparison of $N$(CH) derived for the  17 sources in which both
pairs of transitions {\em viz.,} (3137 \& 3143\,$\AA$) and  (3886 \&
3890\,$\AA$) are available. The straight line shown is fitted to
obtain the ratio between the two column densities.
\label{fig_chcomp}}
\end{figure}

Out of the 24 sources, for 19 sources we have measurements of the
3137 and 3143\,$\AA$ transitions and for 22 sources we have
measurements of the 3886 and 3890\,$\AA$ transitions of CH
(Table\,\ref{tab_meas}). We have re-measured all the CH equivalent
widths from the UVES/VLT archival data and find that in most cases our
measurements are consistent with the published work of
\citet{weselak2010,bhatt2015}. We have revised the values of the
equivalent width of the 3886\,$\AA$ CH line towards HD\,27778 and both
the 3886 and the 3890\,$\AA$ CH lines towards HD\,154445.
Figure\,\ref{fig_chcomp} shows a comparison of CH column densities
derived from a combination of the transitions at 3137 \& 3143\,$\AA$
with $N$(CH) derived from a combination of transitions at 3886 \&
3890\,$\AA$ for the 17 sources in which both sets of CH lines have been
observed.  The values of $N$(CH) show a correlation of 0.99 with
$N$(CH(3137+3143)/$N$(3886+3890) = 1.03$\pm$0.02.  For the rest of the
analysis, we thus use the column densities for the pair of CH
transitions which shows a lower relative uncertainty.

\begin{figure}
\centering
\includegraphics[width=\columnwidth,angle=0]{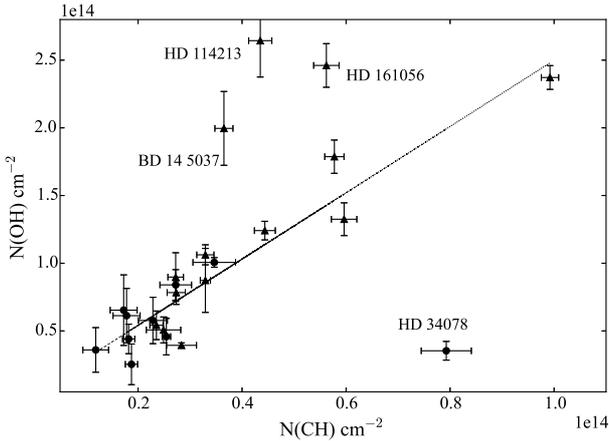}
\caption{Variation of $N$(OH) as a function of $N$(CH).  Filled circles
and triangles denote $N$(CH) derived from the (3137+3143)$\AA$ and
(3886+3890)$\AA$ transitions, respectively. Black dashed line shows the
linear fit to all datapoints excluding the outliers HD\,34078,
HD\,114213, HD\,161056 and BD-14\,5037, which corresponds to a Pearson
correlation coefficient of 0.94. }
\label{fig_ohchcolden}
\end{figure}

Figure\,\ref{fig_ohchcolden} shows a plot of estimated $N$(OH) as a
function of $N$(CH). For clarity, we have used different symbols
to indicate which of the two pairs of transitions was used to derive
the $N$(CH). We find that  for most of the sightlines the column
densities of OH and CH occupy a band in the plot, except towards
BD-14\,5037, HD\,114213, HD\,161056 and HD34078. The overall
correlation coefficient is 0.61 between $N$(OH) and $N$(CH), which is
significantly worse than the value presented by \citet{weselak2009}.
However, the correlation coefficient improves to 0.94 (black dashed
line) when the four outliers are removed, although it is still
less than the 0.99 obtained by \citet{weselak2010}. For the 4 out of 24
sightlines which do not show the correlation between $N$(OH) and
$N$(CH) as seen in the other source, the deviation can not be explained
in terms of the observational uncertainties.

Combined analysis of all available datasets on OH and CH observations
of diffuse clouds enable us to further revise the relation between the
column densities of the two molecules for 20 sightlines to [N(OH)
=  2.41$\pm$0.21 N(CH) +(8.55$\pm$7.17) (in 10$^{12}$\,\cmsq)].
The relationship derived by \citet{weselak2010} had made use of 16
sightlines. The ratio of column densities of OH and CH is found to vary
between 1.5 to 4.1 in diffuse clouds.

We look into the 3 sources HD\,114213, HD\,161056 and BD-14\,5037 for
possible observational reasons behind the very different $N$(CH)-$N$(OH)
relationship, based on the central heliocentric velocities of the
absorption features.  The source BD-14\,5037 shows absorptions at two
velocities around -6\,\kms\ and 6\,\kms. Since only -6\,\kms\ component
is seen in the OH(3081\,$\AA$) spectral line, the numbers we have used
correspond to this component.  Similarly,  the line of sight toward
HD\,114213 shows absorptions at two velocities: CH lines show up at -16
and 6\,\kms, while the OH 3078\,$\AA$ line is at -17\,\kms\ and 3\,\kms\
while the OH(3082\,$\AA$) line is at -20\,\kms\ and 0.3\,\kms.  We have
used here the -20\,\kms, which is identified in all the spectral lines
considered here, since the relative shift in the velocity is smaller
than the observed FWHM of the lines \citep{bhatt2015}.  The sightline
towards HD\,161056 however shows only one velocity component, although
the heliocentric velocities at 3078, 3082, 3137 and 3146\,$\AA$ are
-9.4$\pm$0.1, -11.0$\pm$0.1, -12.7$\pm$0.3 and -14.0$\pm$0.3\,\kms\
respectively \citep{bhatt2015}. As with the other two sources, the
shifts in central velocities are smaller than the FWHM of the lines.
Thus, for the three sources HD\,114213, HD\,161056 and BD-14\,5037 as
with HD\,34078 the lack of correlation between $N$(CH) and $N$(OH) can
not be explained in terms of any observational uncertainties. It is
clearly obvious that along the sightlines towards HD\,114213, HD\,161056
and BD-14\,5037, the OH/CH ratio are 2--3 times the value seen in the
other diffuse clouds.

\section{Molecular hydrogen fraction in diffuse clouds}

Based on theoretical predictions and the observed correlations between
CH and H$_2$, of late $N$(H$_2$)  is often derived from the observed
$N$(CH) assuming CH/H$_2$ = 3.5\,10$^{-8}$ \citep[][and references
therein] {sheffer2008,wiesemeyer2016}. It is also possible to estimate
the column density of hydrogen nuclei using the measured $E(B-V)$ by
using the relationship $N$(H) = 5.8\,10$^{21}$\,$E(B-V)$
\citep{bohlin1978}. For all the lines of sight considered here, since
both $E(B-V)$ and $N$(CH) are available, we have  estimated $f_{\rm
H_2}=\frac{N{\rm (H_2)}}{N{\rm (H)}}$ and do not find any
correlation between $N$(CH)/$N$(OH)) and $f_{\rm H_2}$. The sources,
HD\,161056 and HD\,114213, both of which show high $N$(OH)/$N$(OH)
ratios have $f_{\rm H_2}$ of 51\% and 19\% respectively.  For 17 lines
of sight $f_{\rm H_2}$ is $\leq 0.4$, with BD-14\,5037 and HD\,34078
showing 11\% and 80\% molecular hydrogen fraction, respectively.  We
discuss the anomalously high $N$(CH) shown by HD\,34078 later in the
paper. 

\section{Chemical modeling of the diffuse clouds}

\begin{figure}
\centering
\includegraphics[width=\columnwidth,angle=0]{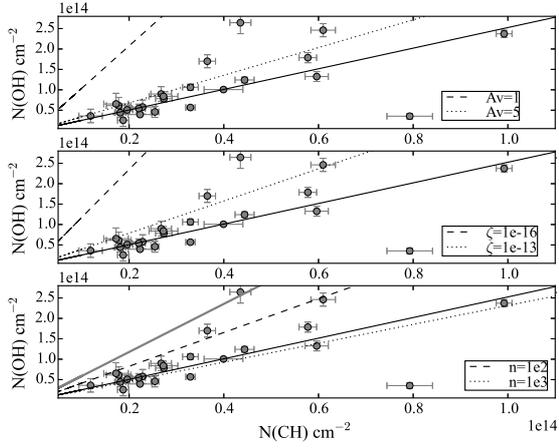}
\caption{Comparison of observed column densities of OH and CH with
computed values from the chemical model Astrochem \citep{maret2015}
used with the chemical network OSU2009. The solid black line in all
panels represents Model $A$ with $n_{\rm H}$=500\,\cmcub, $\zeta_{\rm
CR}$ = 10$^{-14}$\,s$^{-1}$, \av = 2.0, $f_{\rm H_2}$=0.3, $\chi = 5$
and $T_{\rm gas}$=30\,K. {\em Top left}: Model $A$ with \av=1 (dashed) and
\av=5 (dotted); {\em Top Right}: Model $A$ with $\zeta_{\rm CR}$ =
10$^{-16}$\,s$^{-1}$ (dashed) and $\zeta_{\rm CR}$ = 10$^{-13}$\,s$^{-1}$
(dotted); {\em Bottom Left}: Model $A$ with $n_{\rm H}$ = 100\,\cmcub\
(dashed)
and $n_{\rm H}$ = 1000\,\cmcub (dotted). The thick grey line shows the trace of the
model ($n_{\rm H}$=500\,\cmcub, $\zeta_{\rm
CR}$ = 2.5\,10$^{-16}$\,s$^{-1}$, \av = 3.0, $f_{\rm H_2}$=0.3, $\chi = 1$
and $T_{\rm gas}$=30\,K), which explains the OH/CH ratio of HD\,114213.
\label{fig_ohchrat}}
\end{figure}

We have constructed chemical models for the diffuse/translucent clouds
using the Astrochem code \citep{maret2015} together with the reaction
coefficients from the OSU 2009 and examined the predicted OH and CH
abundances.  The models consider a variety of gas phase processes, as
well as simple gas-grain interactions, such as the freeze-out and
desorption via several mechanisms (thermal desorption, cosmic-ray
desorption and photo-desorption).  We have run a grid of models
covering a range of values for the relevant physical parameters in
order to ascertain the parameters to which the CH and OH abundances
are the most sensitive.  We used the abundances of both  CH and OH at
10$^6$\,yr since both attain a constant value beyond 10$^5$\,yr.
Typically in these models, OH abundances are quite sensitive to the
cosmic ray ionization rate ($\zeta_{\rm CR}$), with the abundance
decreasing with increase in $\zeta_{\rm CR}$.
Figure\,\ref{fig_ohchrat} shows a plot of $N$(OH) as a function of
$N$(CH) as predicted by the models which are detailed in the caption.
The purpose of constructing these models is to understand whether
these also predict the  observed correlation between $N$(CH) and
$N$(OH) and not to model the diffuse/translucent clouds each line of
sight accurately. The modeling is also aimed to understand whether the
four lines of sight which lie outside of the correlation band
(Fig.\,\ref{fig_ohchcolden})  are generally consistent with these
simple chemical models that consider gas-grain interactions and
gas-phase reactions but do not include the effects of shock or
turbulence.

We have considered multiple values of $N$(CH) and used the CH
abundances predicted by a model corresponding to a set of input
parameters to estimate $N$(H$_2$) and then used this $N$(H$_2$) to
estimate $N$(OH) from the OH abundances predicted by the model.  In
Fig.\,\ref{fig_ohchrat} we plot the observed values of $N$(OH) and
$N$(CH) for the 24 sightlines presented here.  We find that broadly
the model corresponding to $n_{\rm H}$=500\,\cmcub, $\zeta_{\rm CR}$ =
10$^{-14}$\,s$^{-1}$, \av = 2.0, $f_{\rm H_2}$=0.3, UV radiation
($\chi) = 5$ and $T_{\rm gas}$=30\,K (solid black line in
Fig.\,\ref{fig_ohchrat}) is consistent with many of the observed CH
and OH column densities and we designate this as Model $A$ in the
remaining discussion. For all models discussed here the following
initial abundances relative to H were assumed [He]=0.14, [N]=
2.14\,10$^{-5}$ [O]= 1.76\,10$^{-4}$, [C$^+$]= 7.3\,10$^{-5}$ and
[e$^+$] = 7.3\,10$^{-5}$.

In order to estimate the effect of variation in the input parameters
we also show predictions from models in which the parameters like
$n_{\rm H}$, $\zeta_{\rm CR}$ and \av\ are varied relative to the
parameters of Model $A$, one parameter at a time
(Fig.\,\ref{fig_ohchrat}).  Relative to Model $A$, the presence of
correlation between OH and CH does not change drastically when a)
$f_{\rm H_2}$ is varied up to 0.8, b) $T_{\rm gas}$ is assumed to be
10 and 20\,K and c) $\chi$ is reduced to 1.  In order to show that it
is indeed possible to find a chemical model which is reasonably
consistent with observed parameters and still reproduce the observed
OH/CH ratio for one of the outliers we have constructed a model for
HD\,114213, one of the outliers. Shown in left bottom panel of
Fig.\,\ref{fig_ohchrat} in green continuous line is the prediction of
a model with \av = 3.0 (consistent with observed $E(B-V)$), $n_{\rm
H}$=500\,\cmcub, $\zeta_{\rm CR}$ = 2.5\,10$^{-16}$\,s$^{-1}$, $T_{\rm
gas}$=30\,K and $f_{\rm H_2}$=0.3, which reproduces the observed
$N$(OH) and $N$(CH) quite well. This is by no means a unique model,
since there are several free parameters and fewer observational
constraints, however this shows that the observed column density
ratios are not inconsistent with the chemical models. Thus overall, we
find that with the exception of HD\,34078, the observed ratio of
column densities of OH and CH can be reproduced in all clouds by the
chemical models which do not involve any shock chemistry. The
extremely low $N$(OH) relative to $N$(CH) as seen in HD\,34078 can not
be explained even by varying the input parameters over a much larger
range than what has been shown in Fig.\,\ref{fig_ohchrat}. 

The line of sight towards HD\,34078 has been studied extensively using
FUSE spectra and later millimeter CO observations by
\citet{boisse2005,boisse2009}. This line of sight shows anomalously
high CH/H$_2$ as well as CH$^+$/H$_2$ ratios which have been explained
by the presence of a nascent bow shock around the star, at the
interface between the stellar wind and the ambient interstellar
medium, where the material is strongly compressed \citep{boisse2005}.
There is no corresponding enhancement of OH abundance, which indicates
that the OH absorption is due to the more quiescent H$_2$ gas located
beyond the photodissociation front and shocked region
\citep{boisse2009}. 

The correlation of CH with H$_2$ is explained in terms of the higher
formation rates of CH due to reactions involving molecular hydrogen
\citep{federman1982}. The formation of OH on the other hand is
strongly influenced by grain-surface catalysis, which implies that
presence of dust grains (and hence possibly H$_2$) would enhance OH
formation rates \citep{wagenblast1993}. This can be a possible reason
behind the significant correlation between OH and CH in translucent
clouds. However as seen from the chemical models several combinations
of the input parameters can lead to similar abundance ratios for the
two species. Thus in order to obtain a more quantitative view of the
correlation, accurate modeling of individual sources with further
constraining observations is necessary.

\setlength{\tabcolsep}{0.08cm}
\begin{table*}
\caption{Equivalent widths (from literature and archival data) and 
calculated column densities of OH and CH transitions 
\label{tab_meas}}
\small
\begin{tabular}{llrrrrrrrrrr}
\hline
\hline
\multicolumn{1}{c}{Source} & 
\multicolumn{1}{c}{E$_{B-V}$} & 
\multicolumn{1}{c}{EW(3078)} & 
\multicolumn{1}{c}{EW(3082)} & 
\multicolumn{1}{c}{EW(3137)} & 
\multicolumn{1}{c}{EW(3143)} & 
\multicolumn{1}{c}{EW(3886) }& 
\multicolumn{1}{c}{EW(3890)} & 
\multicolumn{1}{c}{$N$(OH)} & 
\multicolumn{1}{c}{$N$(CH)} & 
\multicolumn{1}{c}{$N$(CH)} \\
\multicolumn{1}{c}{}&
\multicolumn{1}{c}{}&
\multicolumn{1}{c}{}&
\multicolumn{1}{c}{}&  
\multicolumn{1}{c}{}& 
\multicolumn{1}{c}{}&
\multicolumn{1}{c}{}& 
\multicolumn{1}{c}{}&
\multicolumn{1}{c}{}&
\multicolumn{1}{c}{(3137+3143)} & 
\multicolumn{1}{c}{(3886+3890)}\\
\multicolumn{1}{c}{}&
\multicolumn{1}{c}{}&
\multicolumn{1}{c}{m$\AA$}&  
\multicolumn{1}{c}{m$\AA$}& 
\multicolumn{1}{c}{m$\AA$}&
\multicolumn{1}{c}{m$\AA$}& 
\multicolumn{1}{c}{m$\AA$}&
\multicolumn{1}{c}{m$\AA$}&
\multicolumn{1}{c}{\cmsq} & 
\multicolumn{1}{c}{\cmsq} & 
\multicolumn{1}{c}{\cmsq}\\
\hline
HD24398$^1$   & 0.29 &  1.67$\pm$0.08 & 1.11$\pm$0.05 & 3.25$\pm$0.41 & 5.83$\pm$0.36  & 4.96$\pm$0.50 & 3.01$\pm$0.38 & 3.93$\pm$0.18(13) & 2.84$\pm$0.29(13) & 2.23$\pm$0.25(13)\\
HD27778$^{1,3}$   & 0.37 &  5.30$\pm$0.15 & 2.20$\pm$0.10 & 3.15$\pm$1.08 & 9.2$\pm$1.0  & 7.46$\pm$0.60 & 4.84$\pm$0.80 & 1.01$\pm$0.04(14) & 3.366$\pm$0.77(13) & 3.99$\pm$0.41(13)\\
HD34078$^1$   & 0.49 &  1.72$\pm$0.27 & 0.86$\pm$0.21 & 6.92$\pm$1.36 & 19.9$\pm$1.4  & 16.75$\pm$1.30 & 11.28$\pm$0.50 & 3.53$\pm$0.69(13) & 7.34$\pm$0.99(13) & 7.93$\pm$0.48(13)\\
BD-14\,5037$^{a,2}$  & 1.59  & 5.74$\pm$0.43 & 5.71$\pm$0.60 & 3.2$\pm$0.21 & 10.64$\pm$0.33 & 9.47$\pm$0.5 & 6.15$\pm$0.5  & 1.70$\pm$0.16(14) &3.65$\pm$0.17(13) & 4.40$\pm$0.22(13)\\
HD110432$^1$  & 0.48 &  1.81$\pm$0.41 & 1.28$\pm$0.34 &1.83$\pm$0.28 &
5.14$\pm$0.38 &  4.13$\pm$0.20 & 2.40$\pm$0.20 & 4.41$\pm$1.09(13) &
1.92$\pm$0.22(13) &  1.82$\pm$0.12(13)\\
HD114213$^b$  & 1.13 & 9.86$\pm$0.96 & 8.30$\pm$0.87 & 4.48$\pm$0.27 & 10.64$\pm$0.42 & \ldots & \ldots  & 2.64$\pm$0.27(14) & 4.35$\pm$0.22(13) & \ldots \\
HD147889  & 1.08 & 12.19$\pm$0.45 & 5.38$\pm$0.20 & 10.8$\pm$0.24 & 22.48$\pm$0.20 & 20.07$\pm$1.20 & 15.02$\pm$0.98 & 2.37$\pm$0.09(14) & 9.92$\pm$0.17(13) & 1.00$\pm$0.06(14)\\
HD147933$^2$  & 0.48  & 3.63$\pm$0.27 & 2.02$\pm$0.31 & 3.14$\pm$0.28 & 5.69$\pm$0.12 & 5.68$\pm$0.35 & 3.46$\pm$0.26  & 7.83$\pm$0.88(13) & 2.73$\pm$0.17(13) & 2.56$\pm$0.17(13)\\
HD148688$^1$  & 0.55 & 0.92$\pm$0.45 & 0.81$\pm$0.54 & 1.54$\pm$0.22 & 4.28$\pm$0.28  & 3.83$\pm$0.20 & 2.75$\pm$0.20 & 2.53$\pm$1.50(13) & 1.61$\pm$0.17(13) & 1.87$\pm$0.12(13)\\
HD149757  & 0.28 & 2.01$\pm$0.67 & 1.25$\pm$0.32 & 3.54$\pm$0.30 & 6.52$\pm$0.17 & 5.45$\pm$0.20 & 3.57$\pm$0.10 & 4.58$\pm$1.35(13) & 3.10$\pm$0.19(13) & 2.54$\pm$0.08(13)\\
HD151932$^1$  & 0.50 & 4.46$\pm$0.71 & 2.13$\pm$0.54 & 2.68$\pm$0.12 & 7.04$\pm$0.47  & 5.97$\pm$0.78 & 3.60$\pm$0.10 & 8.97$\pm$1.80(13) & 2.72$\pm$0.15(13) & 2.68$\pm$0.22(13)\\
HD152236  &  0.66 & 3.50$\pm$0.44 & 2.41$\pm$0.34 &  \ldots & \ldots & 5.34$\pm$0.45 & 4.15$\pm$0.56 & 8.40$\pm$1.12(13) & \ldots &  2.72$\pm$0.30(13)\\
HD152249  &  0.48 & 2.81$\pm$1.00 & 1.82$\pm$0.80 &\ldots & \ldots &   3.40$\pm$0.50 & 2.61$\pm$0.40 & 6.53$\pm$2.60(13)  &\ldots &  1.72$\pm$0.26(13)\\
HD152270  & 0.50  & 1.95$\pm$0.64 & 2.13$\pm$0.70 & \ldots & \ldots  & 3.52$\pm$0.43 & 2.69$\pm$0.45 & 6.12$\pm$2.01(13) & \ldots & 1.78$\pm$0.26(13)\\
HD154368  & 0.47 & 9.25$\pm$0.58 & 4.01$\pm$0.31 & 5.73$\pm$0.26 & 14.79$\pm$0.24 & 12.24$\pm$1.10 & 9.32$\pm$0.40 & 1.79$\pm$0.12(14) & 5.78$\pm$0.19(13) & 6.18$\pm$0.40(13)\\
HD154445$^{1,3}$  & 0.35 &  2.01$\pm$0.31 & 1.52$\pm$0.33 & 2.43$\pm$0.51 & 6.50$\pm$0.29 &  4.35$\pm$0.50 & 3.76$\pm$0.65 & 5.07$\pm$0.96(13) & 2.49$\pm$0.33(13) & 1.96$\pm$0.12(13)\\
HD154811  & 0.66 &  2.43$\pm$0.70 & 1.64$\pm$0.50 & \ldots & \ldots  & 5.03$\pm$0.34 & 3.12$\pm$0.56 & 5.77$\pm$1.71(13) & \ldots & 2.29$\pm$0.28(13)\\
HD161056$^2$  & 0.59 &  10.71$\pm$0.61 & 6.78$\pm$0.50 & 5.63$\pm$0.30 &
14.23$\pm$0.45 & 10.3$\pm$0.8 & 9.5$\pm$0.55  & 2.46$\pm$0.16(14) & 5.62$\pm$0.24(13) & 5.79$\pm$0.38(13)\\
HD163800  & 0.61 & 2.00$\pm$0.17 & 1.85$\pm$0.12 & 3.42$\pm$0.15 & 7.97$\pm$0.09 & 6.85$\pm$0.30 & 5.16$\pm$0.40 & 5.67$\pm$0.41(13) & 3.29$\pm$0.10(13) & 3.44$\pm$0.21(13)\\
HD164794  & 0.36 & 1.55$\pm$0.40 & 1.00$\pm$0.65 & \ldots & \ldots  & 2.29$\pm$0.54 & 1.83$\pm$0.34 & 3.59$\pm$1.65(13) & \ldots & 1.19$\pm$0.25(13)\\
HD169454  & 1.10 & 6.24$\pm$0.25 & 2.90$\pm$0.22 & 4.52$\pm$0.27 & 11.01$\pm$0.30 & 8.51$\pm$0.54 & 6.47$\pm$0.43 & 1.24$\pm$0.07(14) & 4.44$\pm$0.20(13) & 4.29$\pm$0.28(13)\\
HD170740  & 0.45 & 1.97$\pm$0.43 &  1.73$\pm$0.31 & 2.16$\pm$0.16 & 6.56$\pm$0.22 &  5.10$\pm$0.29 & 2.84$\pm$0.20 & 5.41$\pm$1.06(13) & 2.35$\pm$0.13(13) & 2.20$\pm$0.14(13)\\
HD172028  & 0.79  & 5.05$\pm$0.50 & 4.09$\pm$0.35 & 5.81$\pm$0.35 & 15.59$\pm$0.30 & \ldots & \ldots  & 1.32$\pm$0.12(14) & 5.96$\pm$0.24(13) & \ldots\\
HD210121$^2$  & 0.40  & 4.45$\pm$0.28 & 3.03$\pm$0.23 & 3.31$\pm$0.24 &
8.31$\pm$0.19 & 6.51$\pm$0.31 & 5.83$\pm$0.50  & 1.06$\pm$0.07(14) &
3.29$\pm$0.17(13) & 3.59$\pm$0.25(13)\\
\hline
\end{tabular}
$^a$ BD-14\,5037 shows absorptions at two velocities around -5\,\kms\
and 6\,\kms. Since only -5\,\kms\ component is seen in the
OH(3081\,$\AA$) spectral line, the numbers correspond to this
component.\\
$^b$ HD\,114213 shows absorptions at two velocities around -20\,\kms\
and 6\,\kms\ for CH and -20\,\kms\ and 3\,\kms\ for OH. We present
here the -20\,\kms, which is uniquely identified in all the spectral
lines considered here.\\
$^1$ Sources for which we measured EW(3137) and EW(3143).\\
$^2$ Sources for which we measured EW(3886) and EW(3890).\\
$^3$ Sources for which we derived EW(3886) and EW(3890) slightly
different from the published values.

\end{table*}

\section{Conclusions}

The equivalence of the CH column densities derived from a combination
of transitions at 3137 and 3143\,$\AA$ and a combination of
transitions at 3886 and 3890\,$\AA$ has enabled us to combine types of
CH datasets. This has resulted in an increase in the number of
sightlines for which both $N$(OH) and $N$(CH) have been observed.
With the exception of four sightlines, the column density of OH
appears to correlate well with the CH column density in all sources.
The derived relationship between $N$(OH) and $N$(CH) is thus based on
a dataset with larger number of sightlines observed with lower
uncertainties than the previous determinations. These results also
verify the oscillator strengths for all the transitions which have
been observed. We find that the $N$(OH)/$N$(CH) ratio is not
correlated with the fraction of molecular hydrogen in these clouds. We
show that except for the line of sight toward HD\,34078 which is
thought to show signatures of shock-induced increase in CH abundance,
the observed OH/CH ratios in all other clouds are  well reproduced by
the chemical models considered here.

\section*{Acknowledgements}
BM sincerely thanks the referee Jacek Krelowski for suggestions which
enhanced the scope of the paper by including additional data. BM thanks
J. P. Ninan for his help with many python related issues.  This
research has made use of the VizieR catalogue access tool, CDS,
Strasbourg, France. The original description of the VizieR service was
published in A\&AS 143, 23.

\bsp    
\label{lastpage}

\begin{thebibliography}{99}
\bibitem[\protect\citeauthoryear{Bhatt \& Cami}{2015}]{bhatt2015} 
Bhatt N.~H., Cami J., 2015, ApJS, 216, 22
\bibitem[\protect\citeauthoryear{Bohlin, Savage, 
\& Drake}{1978}]{bohlin1978} Bohlin R.~C., Savage B.~D., Drake J.~F., 1978, ApJ, 224, 132
\bibitem[\protect\citeauthoryear{Boiss{\'e} et al.}{2009}]{boisse2009}
Boiss{\'e} P., et al., 2009, A\&A, 501, 221 
\bibitem[\protect\citeauthoryear{Boiss{\'e} et al.}{2005}]{boisse2005}
Boiss{\'e} P., Le Petit F., Rollinde E., Roueff E., Pineau des
For{\^e}ts G., Andersson B.-G., Gry C., Felenbok P., 2005, A\&A, 429,
509 
\bibitem[\protect\citeauthoryear{Chaffee 
\& Lutz}{1977}]{chaffee1977} Chaffee F.~H., Jr., Lutz B.~L., 1977, ApJ, 213, 394 
\bibitem[\protect\citeauthoryear{Crutcher \&
Watson}{1976}]{crutcher1976} Crutcher R.~M., Watson W.~D., 1976, ApJ,
203, L123 
\bibitem[\protect\citeauthoryear{Federman}{1982}]{federman1982}
Federman S.~R., 1982, ApJ, 257, 125 
\bibitem[\protect\citeauthoryear{Felenbok \& Roueff}{1996}]{felenbok1996} 
Felenbok P., Roueff E., 1996, ApJ, 465, L57 
\bibitem[\protect\citeauthoryear{Herbig}{1968}]{herbig1968} 
Herbig G.~H., 1968, ZA, 68, 243
\bibitem[\protect\citeauthoryear{Kre{\l}owski et
al.}{1999}]{krelowski1999} Kre{\l}owski J., Ehrenfreund P., Foing
B.~H., Snow T.~P., Weselak T., Tuairisg S.~{\'O}., Galazutdinov G.~A.,
Musaev F.~A., 1999, A\&A, 347, 235 
\bibitem[\protect\citeauthoryear{Lien}{1984}]{lien1984} Lien 
D.~J., 1984, ApJ, 284, 578
\bibitem[\protect\citeauthoryear{Maret 
\& Bergin}{2015}]{maret2015} Maret S., Bergin E.~A., 2015, ascl.soft, ascl:1507.010
\bibitem[\protect\citeauthoryear{Mattila}{1986}]{mattila1986} Mattila
K., 1986, A\&A, 160, 157 
\bibitem[\protect\citeauthoryear{Roueff}{1996}]{roueff1996} Roueff 
E., 1996, MNRAS, 279, L37 
\bibitem[\protect\citeauthoryear{Sheffer et al.}{2008}]{sheffer2008}
Sheffer Y., Rogers M., Federman S.~R., Abel N.~P., Gredel R., Lambert
D.~L., Shaw G., 2008, ApJ, 687, 1075 
\bibitem[\protect\citeauthoryear{Wagenblast et
al.}{1993}]{wagenblast1993} Wagenblast R., Williams D.~A., Millar
T.~J., Nejad L.~A.~M., 1993, MNRAS, 260, 420
\bibitem[\protect\citeauthoryear{Weinreb et 
al.}{1963}]{weinreb1963} Weinreb S., Barrett A.~H., Meeks M.~L., 
Henry J.~C., 1963, Natur, 200, 829 
\bibitem[\protect\citeauthoryear{Weselak et 
al.}{2010}]{weselak2010} Weselak T., Galazutdinov G.~A., Beletsky 
Y., Kre{\l}owski J., 2010, MNRAS, 402, 1991 
\bibitem[\protect\citeauthoryear{Weselak et al.}{2009}]{weselak2009} 
Weselak T., Galazutdinov G., Beletsky Y., Kre{\l}owski J., 2009, A\&A, 499, 783 
\bibitem[\protect\citeauthoryear{Weselak et al.}{2004}]{weselak2004} 
Weselak T., Galazutdinov G.~A., Musaev F.~A., Kre{\l}owski J., 2004, A\&A, 414, 949
\bibitem[\protect\citeauthoryear{Wiesemeyer et 
al.}{2016}]{wiesemeyer2016} Wiesemeyer H., et al., 2016, A\&A, 585, A76 
\end{thebibliography}
\end{document}